\documentstyle[12pt,epsfig,amstex,amssymb]{article}
\newcommand{\av}[1]{\mbox{$ \langle #1 \rangle $}}
\newcommand{\lsim}{\raisebox{-0.5mm}{$\stackrel{<}{\scriptstyle{\sim}}$}}

\newcommand{\qsq}{\mbox{$Q^2$}}
\newcommand{\s}{\mbox{$s$}}
\newcommand{\sn}{\mbox{$s_n$}}
\newcommand{\x}{\mbox{$x$}}
\newcommand{\y}{\mbox{$y$}}
\newcommand{\R}{\mbox{$R$}}
\newcommand{\Ee}{\mbox{$E_e^0$}}
\newcommand{\Ep}{\mbox{$E_p^0$}}
\newcommand{\Eg}{\mbox{$E_\gamma$}}
\newcommand{\FL}{\mbox{$F_L(x,Q^2)$}}
\newcommand{\rh}{\mbox{$\rho(R)$}}
\newcommand{\rhR}{\mbox{$\rho(R;\varepsilon_0)$}}
\newcommand{\rhRz}{\mbox{$\rho(R=0;\varepsilon_0)$}}
\newcommand{\rhRi}{\mbox{$\rho(R=\infty;\varepsilon_0)$}}
\newcommand{\srhR}{\mbox{$\sigma_{\rho,R}$}}

\newcommand{\srhRzsq}{\mbox{$\sigma^2_{\rho,R=0}$}}

\newcommand{\srhRisq}{\mbox{$\sigma^2_{\rho,R=\infty}$}}
\newcommand{\gev}{\mbox{\rm GeV}}

\newcommand{\gevsq}{\mbox{${\rm GeV}^2$}}

\newcommand{\pbinv}{\mbox{${\rm pb^{-1}}$}}
\begin{document}
\begin{flushright}
LAL 96-32\\
IIHE 96-01\\
\end{flushright}
\vspace*{2.5cm}
\centerline {\LARGE\bf On the possibility of measuring $F_L(x,Q^2)$}
\vspace{2mm}
\centerline {\LARGE\bf  at HERA using radiative events}
\vspace{1cm}
\centerline {\sc L. Favart$^{a}$, M. Gruw\'e$^{b}$, P. Marage}
\vspace{0.5cm}
\centerline {\it Universit\'e Libre de Bruxelles, B-1050 Brussels, Belgium}
\vspace{0.5cm}
\centerline {\sc Z. Zhang}
\vspace{0.5cm}
\centerline {\it Laboratoire de l'Acc\'el\'erateur Lin\'eaire, IN2P3 - CNRS and}
\centerline {\it Universit\'e de Paris-Sud, F-91405, Orsay Cedex, France}
\vspace{1cm}
\centerline {$^{a}$ Now at Laboratoire de l'Acc\'el\'erateur Lin\'eaire, Orsay, France}
\centerline {$^{b}$ Now Fellow at CERN, Geneva, Switzerland}
\vspace{3cm}
\centerline {\large Abstract}
\bigskip
\begin{center}
\begin{minipage}{14.5cm}
{It is shown that a significant measurement of the longitudinal structure function 
$F_L(x,Q^2)$ can be performed at HERA, for $\qsq = 2$ \gevsq\ and $\qsq = 5$ \gevsq\
and for \x\ around $10^{-4}$, 
using radiative events with hard photon emission collinear to the incident lepton beam, 
under the present running conditions and with an integrated luminosity of 10 \pbinv.
The influence of experimental conditions is discussed.}
\end{minipage}
\end{center}
\newpage

\section{Introduction}

The measurement of the longitudinal structure function of the proton, \FL, is an
important task at the $HERA$ $ep$ collider. 
The knowledge of $F_L$ is needed to extract in a model independent way 
the structure function $F_2$ from the measured cross section. Moreover, 
the measurement of the \qsq\ dependence of $F_L$ allows QCD tests, and the measurement
of its \x\ dependence makes possible to constrain the gluon distribution function in 
the proton \cite{Cooper}.

For measuring \FL, it is necessary to vary the $ep$ centre of mass energy.
This can be achieved by running the collider with reduced beam energy \cite{Cooper},
but this procedure has the obvious draw back that a significant running time is lost for 
high energy physics and that the collider is not operated in optimal conditions.
For the $F_L$ measurement itself, a major experimental problem is the photoproduction
background, when a hadron is wrongly taken as the electron candidate. 
Another important source of systematic error is the relative 
normalisation of data sets obtained in different beam conditions.
This difficulty explains why measurements of \FL\ are so delicate, and why only a few 
fixed target results have been published \cite{ref_exp_fl}.  

Another method has been proposed by Krasny et al. \cite{Krasny}. It makes use of
deep inelastic radiative events in which a real photon has been emitted in the
direction of the incident electron beam, which corresponds to an effective decrease 
of the beam energy.
The advantages of this method are that it can be used in parallel with normal data
taking, that it avoids luminosity normalisation problems, and that the statistical 
and systematic precisions increase continuously during data taking.

In this letter, it is shown that, using a slightly modified procedure, this 
latter method 
allows performing a first significant measurement of $F_L$ with the present collider 
and detector conditions.


\section{Experimental Procedure}

The differential cross section for deep inelatistic $ep$ scattering can be written\footnote{
Contributions due to weak boson exchange are negligible in the $Q^2 <\!< M^2_{Z^0}$ domain 
relevant for this letter.} 
as

\begin{eqnarray}
 \frac{d^2 \sigma}{dxdQ^2} 
 & = & \frac{4\pi \alpha^2}{xQ^4} \ \left[y^2xF_1(x,Q^2)+(1-y)F_2(x,Q^2)\right]
 \nonumber \\
 & = & \frac{\alpha}{2\pi xQ^2} \ \left[1+(1-y)^2\right] \ \left[1+\varepsilon
  R \right] \ \sigma_T\,, \label{eq:xsec}
\end{eqnarray}
where $F_1(x,Q^2)$ and $F_2(x,Q^2)$ are the proton structure functions,  
and $R(x,Q^2)$ is related to the longitudinal structure function
$F_L(x,Q^2)= F_2(x,Q^2) - 2xF_1(x,Q^2)$ by
$R(x,Q^2)=F_L(x,Q^2)/2xF_1(x,Q^2)=\sigma_L / \sigma_T$, 
$\sigma_L$ and $\sigma_T$ being the cross sections for the scattering
of transversely and longitudinally polarized virtual photons;
$\varepsilon$ is the polarisation parameter
\begin{equation}
\varepsilon=\frac{2 \ (1-y)}{1+(1-y)^2}.
              \label{eq:varepsilon}
\end{equation}
The kinematical variables
are $\qsq= -q^2$
and the two Bjorken variables $ x= \qsq  / 2p \cdot q $ 
and $ y= p \cdot q / p \cdot k$, where $k$, $p$, 
$q$ are, respectively, the four-momenta of the incident electron, of the incident proton 
and of the virtual photon.
These variables obey the relation 
\begin{equation}
\y = \qsq / x \cdot s , 
              \label{eq:y}
\end{equation}
where\footnote{The electron and proton masses are neglected.}
$s = 2 \ k \cdot p \simeq 4 \Ee \Ep $ is the square of the $ep$ centre of mass energy, 
$\Ee\ = 27.5\ \gev$ and $\Ep\ = 820\ \gev$ being respectively the incident electron and 
proton beam energies.

The basic principle of the $R(x,\qsq)$ measurement is to perform a linear fit of cross 
section (\ref{eq:xsec}) as a function of $\varepsilon$.
As $\varepsilon$ depends on \y, which in turn is related to \qsq\ and \x\ through relation 
(\ref{eq:y}), varying $\varepsilon$ implies varying \s, i.e. the beam energies.

Because of the good angular separation between the incident and the scattered electron 
directions, deep inelastic events $e + p \rightarrow e + X + \gamma$ containing a 
photon of energy \Eg\ emitted in the direction of the incident electron beam  
can be interpreted as interactions with an incident electron energy effectively reduced by the 
factor
\begin{equation}
z=\frac{\Ee -\Eg}{\Ee}.
              \label{eq:z}  
\end{equation}
The square of the $ep$ centre of mass energy, \s, is reduced by the
same factor with respect to the nominal value \sn, obtained using for the
incident electron the beam energy, 27.5 \gev.

For the radiative events, the kinematical variables are computed with the
reduced electron energy. Relations (\ref{eq:xsec}) to (\ref{eq:y}) thus apply 
unchanged to this sample.
The spectrum of measured photon energies induces, for given \x\ and \qsq\
values, a continuous distribution of the \y, \s\ and $\varepsilon$ variables. 
This is made explicit in the following relation, where $\varepsilon$ is computed in
function of $z$ and \sn :
\begin{equation}
\varepsilon=z\sn \frac{2x(xz\sn-\qsq)}{x^2z^2\sn^2 + (xz\sn-\qsq)^2}.
              \label{eq:epsil}  
\end{equation}

The original idea in \cite{Krasny} was to use radiative events and to measure \R\ by 
fitting the slope of the 
$\varepsilon$ dependence of cross section (\ref{eq:xsec}).
This procedure is independent of the knowledge of $F_2$, but it requires a very large 
integrated luminosity (of the order of 200 \pbinv) to provide a significant measurement 
of \R.
We propose instead to use the available measurements of $F_2$, and to exploit the
dependence on \R\ of the {\it shape} of the $\varepsilon$ distribution. 

\begin{figure}[htbp]
\begin{center}
\begin{picture}(50,410)
\put(-225,-210){
\epsfig{file=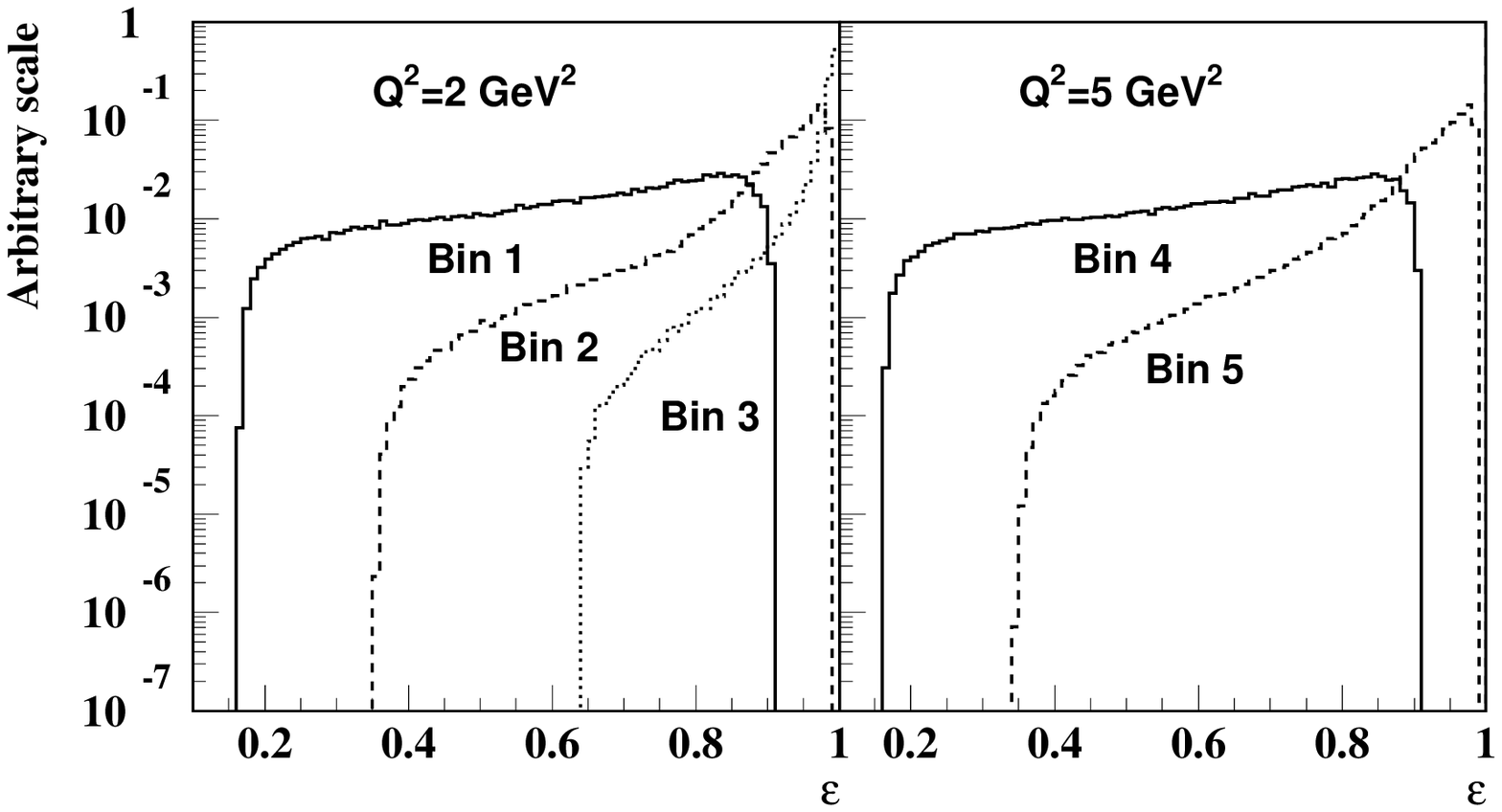,height=180mm,width=180mm}}
\end{picture}
\caption{Distributions of the $\varepsilon$ parameter in the selected bins,
for the GRV parameterisation of the $F_2$ structure function and $R = 0.5$.}
\label{fig:eps_4bins}
\end{center}
\end{figure}
\begin{figure}[htbp]
\begin{center}
\begin{picture}(50,410)
 \put(-205,10){
\epsfig{file=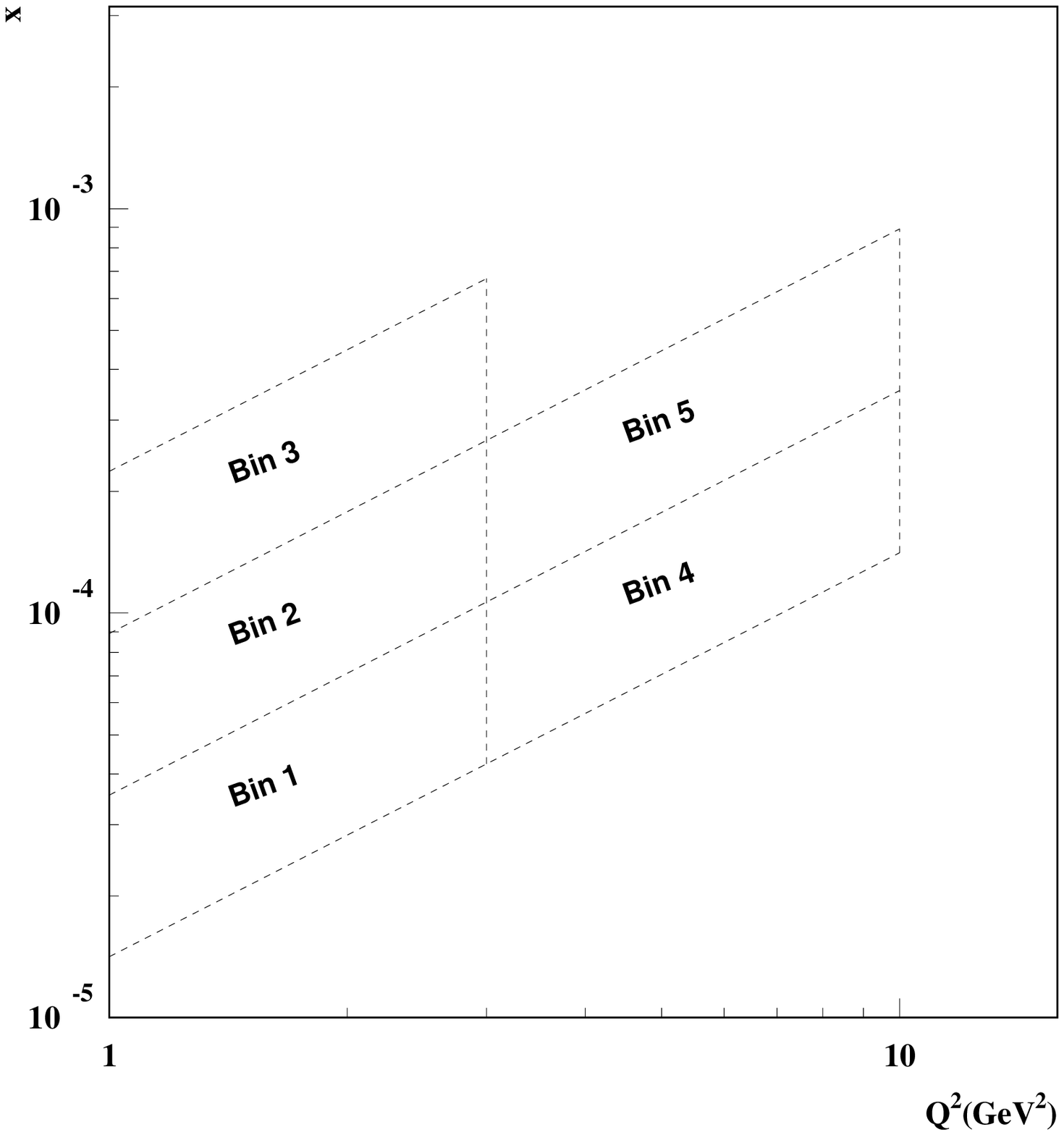,height=170mm,width=170mm}}
\end{picture}
\caption{Selected (\x,\qsq) bins.}
\label{fig:flbins}
\end{center}
\end{figure}

The  $\varepsilon$ distributions are presented in Fig. \ref{fig:eps_4bins} for 
the 5 bins in \x\ and \qsq\ shown in Fig. \ref{fig:flbins}.
Three bins are designed for \av{\qsq} = 2 \gevsq\ (with \av{\x} ranging from $4 \cdot 10^{-5}$
to $2 \cdot 10^{-4}$), and two bins for \av{\qsq} = 5 \gevsq\ (with \av{\x} ranging from 
$10^{-4}$ to $3 \cdot 10^{-4}$).
The $\varepsilon$ distributions are obtained from a Monte Carlo simulation using the GRV 
\cite{GRV} parameterisation of the $F_2$ proton structure function.

The simulated integrated luminosity corresponds to 10 \pbinv, which is the luminosity 
delivered by HERA in 1995, and the following kinematical cuts are applied:
\begin{eqnarray}
& & E_{e^{\prime}}>2\, {\rm GeV}\,, \nonumber \\
& & \theta_{e^{\prime}}<177^{\circ}\,, \nonumber \\
& & E_{\gamma}>4\, {\rm GeV}\, , 
              \label{eq:cuts}  
\end{eqnarray}
where $E_{e^{\prime}}$ and $\theta_{e^{\prime}}$ are the energy and the polar angle 
(defined with respect to the proton beam direction) of the scattered electron.
These are realistic hypotheses for the present detector and trigger conditions.

The \R\ dependence of the $\varepsilon$ distribution can be studied using the variable 
$\rhR$, defined as the ratio of the numbers of events with $\varepsilon$ smaller or 
larger than a chosen value $\varepsilon_0$:
\begin{equation}
\rhR=\frac{N(R;\varepsilon<\varepsilon_0)}{N(R;\varepsilon>\varepsilon_0)}.
              \label{eq:rho}  
\end{equation}

\begin{figure}[htbp]
\begin{center}
\begin{picture}(70,510)
\put(-195,-10){
\epsfig{file=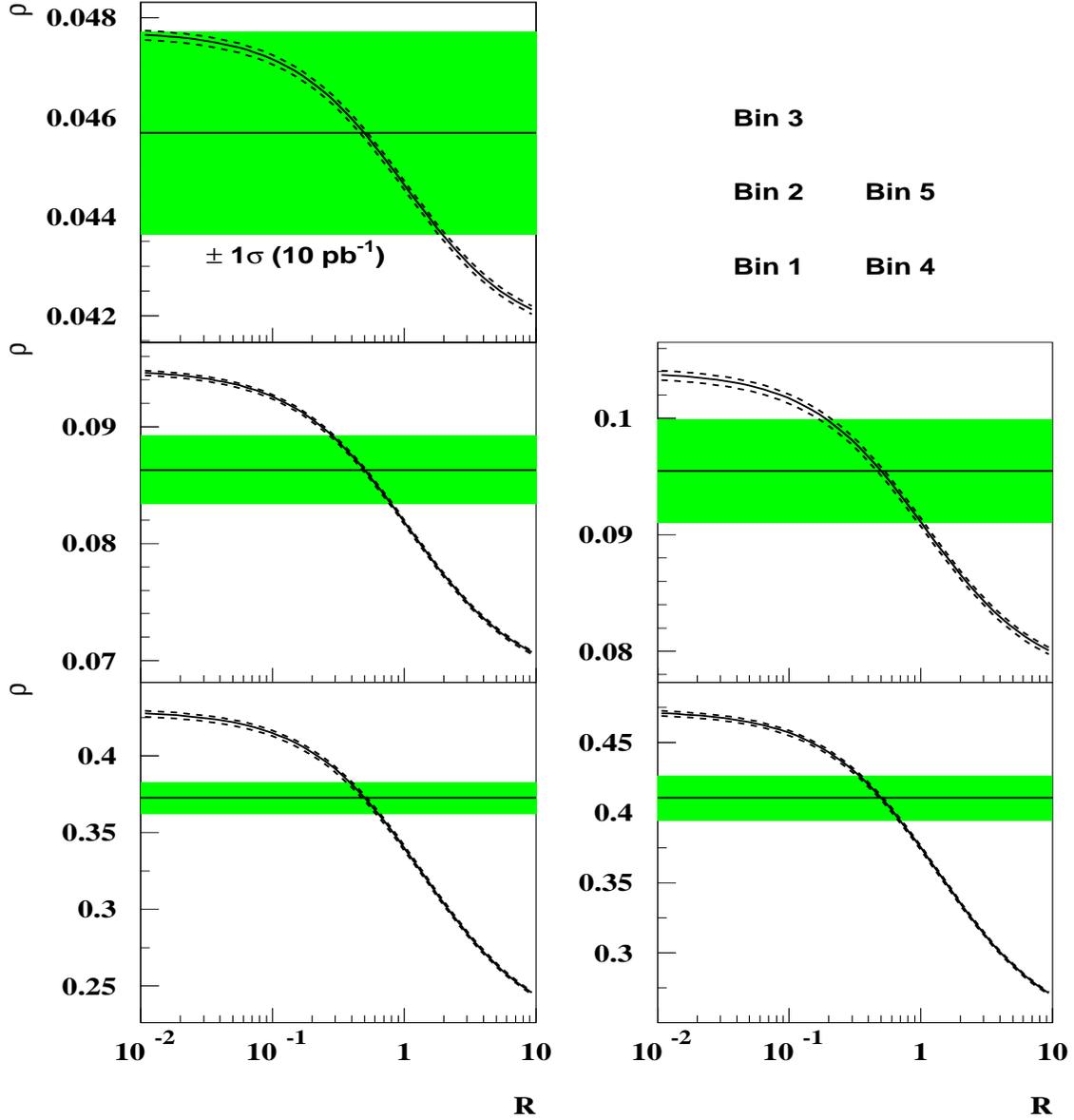,height=170mm,width=170mm}}
\end{picture}
\end{center}
\caption{\rh\ dependence on \R\ for the GRV parameterisation 
(solid curves) and for the modified parameterisations described in the text 
(dashed curves), in the selected bins. 
The grey bands correspond to $\pm 1\sigma$ statistical errors for an
integrated luminosity of 10 \pbinv\ and for $R = 0.5$.}
\label{fig:rho_4bins}
\end{figure}

Fig. \ref{fig:rho_4bins} shows the \rhR\ dependence on $R$ in the selected (\x,\qsq) 
bins, for the input typical value $R = 0.5$ as obtained in a recent H1 analysis 
\cite{H1_FL}.
Since each bin covers a different $\varepsilon$ range,
the chosen optimal $\varepsilon_0$ value is bin dependent.
The dashed curves show the \rh\ distribution for an input $F_2$ structure function
modified by $\pm 10 \%$ at $x = 10^{-4}$, the modification decreasing linearly 
to $\pm 5\%$ at $x = 10^{-2}$. 
This corresponds to a conservative estimate of the present uncertainty on $F_2$.
The grey bands correspond to the statistical precision of the \rh\ measurement for 
an integrated luminosity of 10 \pbinv.

\begin{figure}[htbp]
\begin{center}
\begin{picture}(50,410)
\put(-225,-210){
\epsfig{file=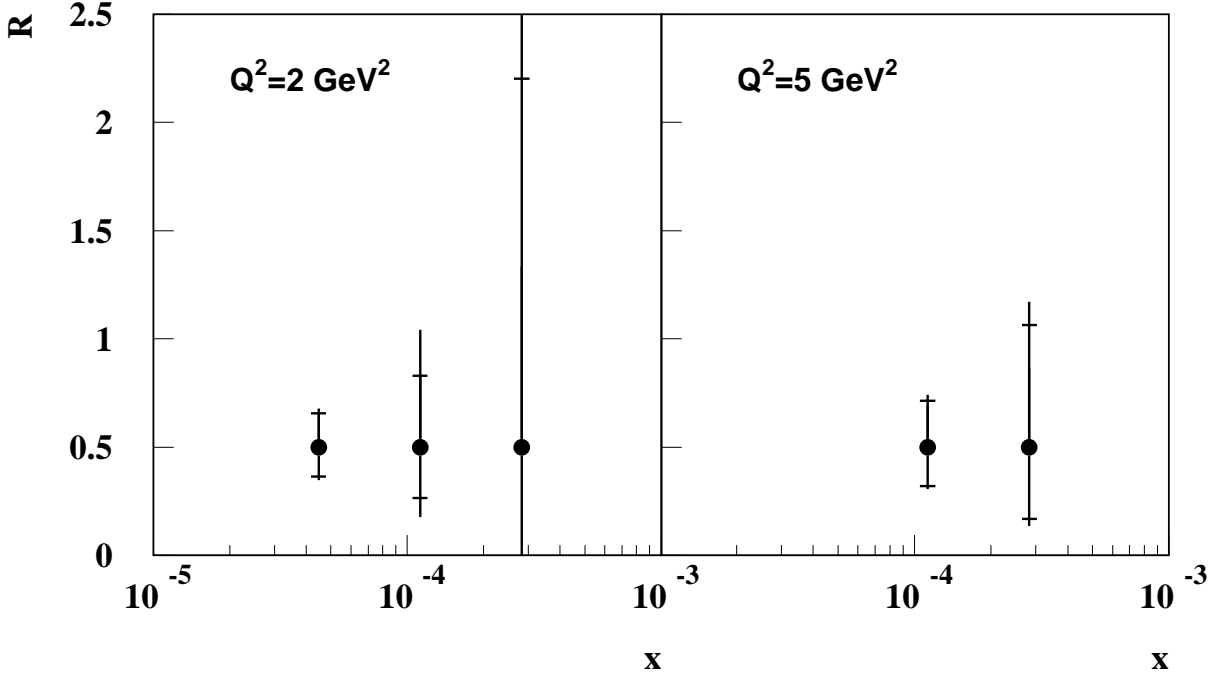,height=180mm,width=180mm}}
\end{picture}
\end{center}
\caption{Typical precision of the \R\ measurement in the selected bins, under
the experimental conditions specified in eq. (\ref{eq:cuts}). 
The inner error bars show the measurement precision for an
integrated luminosity of 10 \pbinv, taking also into account the uncertainty
on the structure function parameterisation described in the text.
The outer error bars include, added in quadrature, the effects of the uncertainties
on $E_{e^{\prime}}$, $\theta_{e^{\prime}}$ and $E_{\gamma}$. }
\label{fig:R_4bins}
\end{figure}

The measurement of \R\ is deduced from the intersection of the grey bands with
the spread of curves describing the \rh\ dependence of the input structure 
function $F_2$. The inner error bars in Fig. \ref{fig:R_4bins} show the statistical 
precision of the \R\ measurement 
for the cuts (\ref{eq:cuts}), an integrated luminosity of 10 \pbinv\ and the quoted 
uncertainty on $F_2$.

\section{Discussion and Conclusions}\label{discussion}

To estimate the sensitivity of the proposed method to several experimental 
parameters, the variable $\Sigma$ is defined as
\begin{equation}
\Sigma(\varepsilon_0,{\cal L},F_2)=\frac{|\rhRz-\rhRi|}
                      {\sqrt{\srhRzsq+\srhRisq}}
              \label{eq:Sigma}  
\end{equation}
for a given choice of $\varepsilon_0$, of the integrated luminosity ${\cal L}$
and of the input $F_2$ structure function, \srhR\ being the statistical error 
on \rh, estimated through the Monte Carlo simulation.
This variable quantifies the possibility of distinguishing between 
the two extreme values of \R: $R = 0$ and $R = \infty$. 

\begin{figure}[htbp]
\begin{center}
\begin{picture}(50,410)
\put(-225,-210){
\epsfig{file=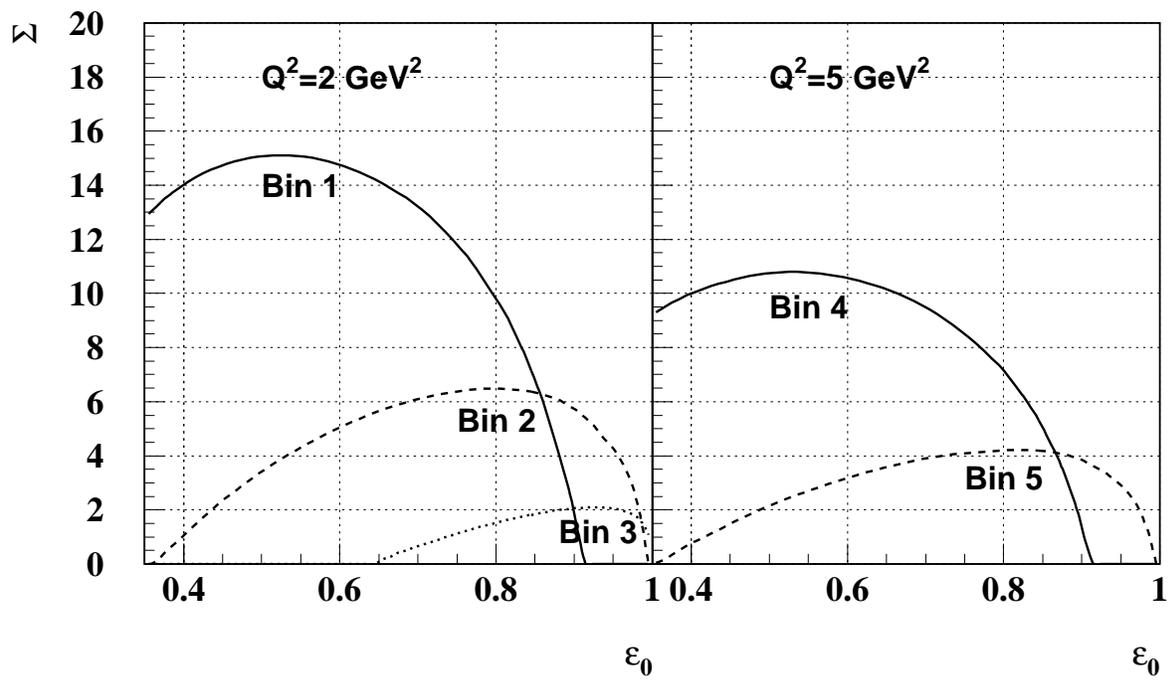,height=180mm,width=180mm}}
\end{picture}
\end{center}
\caption{Dependence of the sensitivity $\Sigma$ 
 on the $\varepsilon$ parameter in the selected bins, 
for the GRV structure function.}
\label{fig:epsil0}
\end{figure}

Fig. \ref{fig:epsil0} shows that the sensitivity $\Sigma$ for each 
$(x,Q^2)$ bin is only weakly dependent on the 
$\varepsilon_0$ value over a rather large domain in $\varepsilon_0$. 
It is found that it also depends little on detector smearing effects.

\begin{figure}[htbp]
\begin{center}
\begin{picture}(70,510)
\put(-195,-10){
\epsfig{file=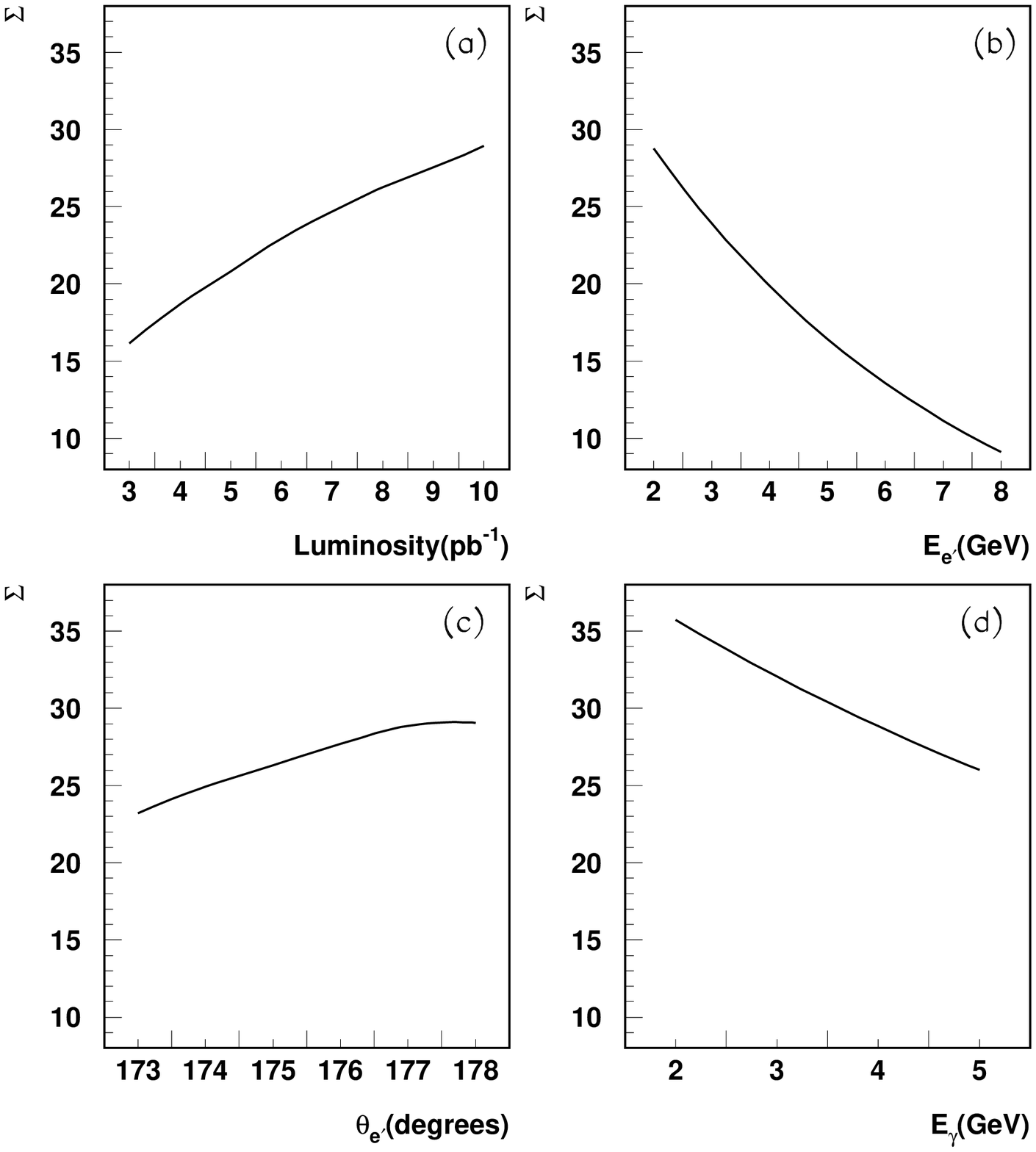,height=160mm,width=160mm}}
\end{picture}
\end{center}
\caption{Sensitivity $\Sigma$ as a function of
a) the integrated luminosity;
b)$E_{e^{\prime}}$;
c) $\theta_{e^{\prime}}$; 
d) $E_{\gamma}$,
for the GRV parameterisation
under the experimental conditions specified in eq. (\ref{eq:cuts})
and in the total kinematical region covered by the selected bins.}
\label{fig:sensit}
\end{figure}

On the other hand, as can be seen in Fig. \ref{fig:sensit}, the sensitivity $\Sigma$ is 
strongly dependent on the detector acceptance conditions, in particular the electron energy 
threshold which is related to the $y$ and $\varepsilon$ ranges.
For the same luminosity, the sensitivity is enhanced by a factor 2.1 for $E_{e^{\prime}}$ 
decreasing from 6 to 2 \gev.
A decrease on the photon energy threshold $E_{\gamma}$ also improves significantly the 
sensitivity.
The lowering of the electron energy threshold is a challenge for the HERA experiments
because of the significant background from photoproduction interactions in which
low energy hadrons are misidentified as the scattered electron.

Studies have been performed of the effects of experimental uncertainties,
which are in general bin to bin dependent.
The detector resolution was simulated using realistic smearing functions for
the H1 experiment; in addition, systematic uncertainties were taken into account
(1\% on $E_{e^{\prime}}$, 1 mrad on $\theta_{e^{\prime}}$ and 1.5\% on $E_{\gamma}$).
The effects of these uncertainties, combined in quadrature with the statistical
errors and the effects of the uncertainty on the $F_2$ structure function, are 
displayed as the outer error bars on the \R\ measurements of Fig. \ref{fig:R_4bins}.

The subtraction of the remaining photoproduction background is another important
source of systematic uncertainty, which affects mostly the lower x bins. 
There, it is found to induce systematic errors of the same order as the errors due 
to detector resolution.
Another source of systematic error will be the overlap of non radiative deep 
inelastic events with bremsstrahlung events for which the photon is detected in the
photon detector and the scattered electron is not detected.
An electron tagger with a large energy acceptance is an important tool to reduce 
this background.
As for the present uncertainty on the $F_2$ structure function, it is observed in 
Fig. \ref{fig:rho_4bins} that it does not imply a large systematic uncertainty on \R.

Taking all these effects into account, it is found that for an integrated
luminosity of 10 \pbinv\ the 
statistical errors dominate over the systematic errors in most of the chosen bins. 
With increased statistics, a significant improvement of the measurement
precision is thus to be expected.
Detailed opimisation studies are also expected to improve the measurement precision.
 
In summary, it is possible to measure at HERA the proton longitudinal 
structure function $F_L(x,Q^2)$ without changing the nominal beam energies, 
using deep inelastic radiative events for which a hard photon is emitted collinear 
to the incident lepton, which results in an effective reduction of the incident
electron energy. 
In the present running conditions, and with an integrated luminosity
of about 10 \pbinv, a first significant measurement of $R$ is possible
for \qsq\ \lsim\ 5\ \gevsq\ and \x\ around $10^{-4}$.
The precision of the measurement will improve continuously as more luminosity
is delivered by HERA.

\section*{Acknowledgements}
It is a pleasure to thank our colleagues in H1 for interest and support in the
present studies.

\newpage

\begin{thebibliography}{99}
\bibitem{Cooper} 
  A.M. Cooper-Sarker et al., in Proc. of the HERA Workshop, 
  Hamburg 1987, ed. R.D. Peccei, Vol.1, p. 231; \\
  A.M. Cooper-Sarker et al., in Proc. of the HERA Workshop, 
  Hamburg 1991, eds. W. Buchm\"{u}ller and G. Ingelman, Vol.1, p. 155.
\bibitem{ref_exp_fl}
  J.J. Aubert et al., Nucl.Phys. {\bf B259} (1985) 109; {\bf B293} (1987) 740;\\
  A.C. Benvenuti et al., Phys. Lett. {\bf B223} (1989) 485; {\bf B237} (1990) 592;\\
  L.W. Whithlow, PhD thesis, SLAC-report-357 (1990) and references therein;\\
  L.W. Whithlow et al., Phys. Lett. {\bf B250} (1990) 193.;\\ 
  J.P. Berge et al., Z. Phys. {\bf C49} (1991) 187.
\bibitem{Krasny} 
  M.W. Krasny et al., Z. Phys. {\bf C53} (1992) 687;\\
  W. P{\l}aczek, Ph.D. thesis, Acta Physica Polonica {\bf B24} (1993) 1229.
\bibitem{GRV} 
  M. Gl\"uck, E. Reya and A. Vogt, DESY 94-206.
\bibitem{H1_FL}
  H1 Coll., Int. Workshop on Deep Inelastic Scattering and Related Phenomena, 
  Rome 1996, to be publ. 

\end{thebibliography}
\end{document}